\documentclass{resonance}
\usepackage{amsfonts}
\usepackage{amsmath}
\usepackage{amssymb}
\usepackage{rotating}
\usepackage{epsfig}  
\usepackage{graphics} 
\usepackage{graphicx} 
\usepackage{txfonts}
\usepackage{marginnote}
\usepackage{array}
\usepackage{geometry}
\usepackage{color}      
\definecolor{light}{rgb}{0.8,0.5,0.5}
\definecolor{cblue}{rgb}{0.9,0.9,1.0}
\definecolor{darkblue}{rgb}{0.1,0.1,0.6}
\definecolor{darkred}{rgb}{0.6,0.1,0.1}
\usepackage{cleveref}
\crefformat{section}{\S#2#1#3}
\crefformat{subsection}{\S#2#1#3}
\crefformat{subsubsection}{\S#2#1#3}
\crefrangeformat{section}{\S\S#3#1#4 to~#5#2#6}
\crefmultiformat{section}{\S\S#2#1#3}{ and~#2#1#3}{, #2#1#3}{ and~#2#1#3}
\newcommand{\bed}{\begin{displaymath}}
\newcommand{\eed}{\end{displaymath}}
\newcommand{\bei}{\begin{itemize}}
\newcommand{\eei}{\end{itemize}}
\newcommand{\bef}{\begin{figure}}
\newcommand{\eef}{\end{figure}}
\newcommand{\ben}{\begin{enumerate}}
\newcommand{\een}{\end{enumerate}}
\newcommand{\beq}{\begin{equation}}
\newcommand{\eeq}{\end{equation}}
\newcommand{\ber}{\begin{eqnarray}}
\newcommand{\eer}{\end{eqnarray}}

\newcounter{attnctr} 

\begin{document}

\title{The Sounds of Music : Science of Musical Scales}
\secondTitle{III : Indian Classical}
\author{Sushan Konar}

\maketitle
\authorIntro{\includegraphics[width=2.5cm]{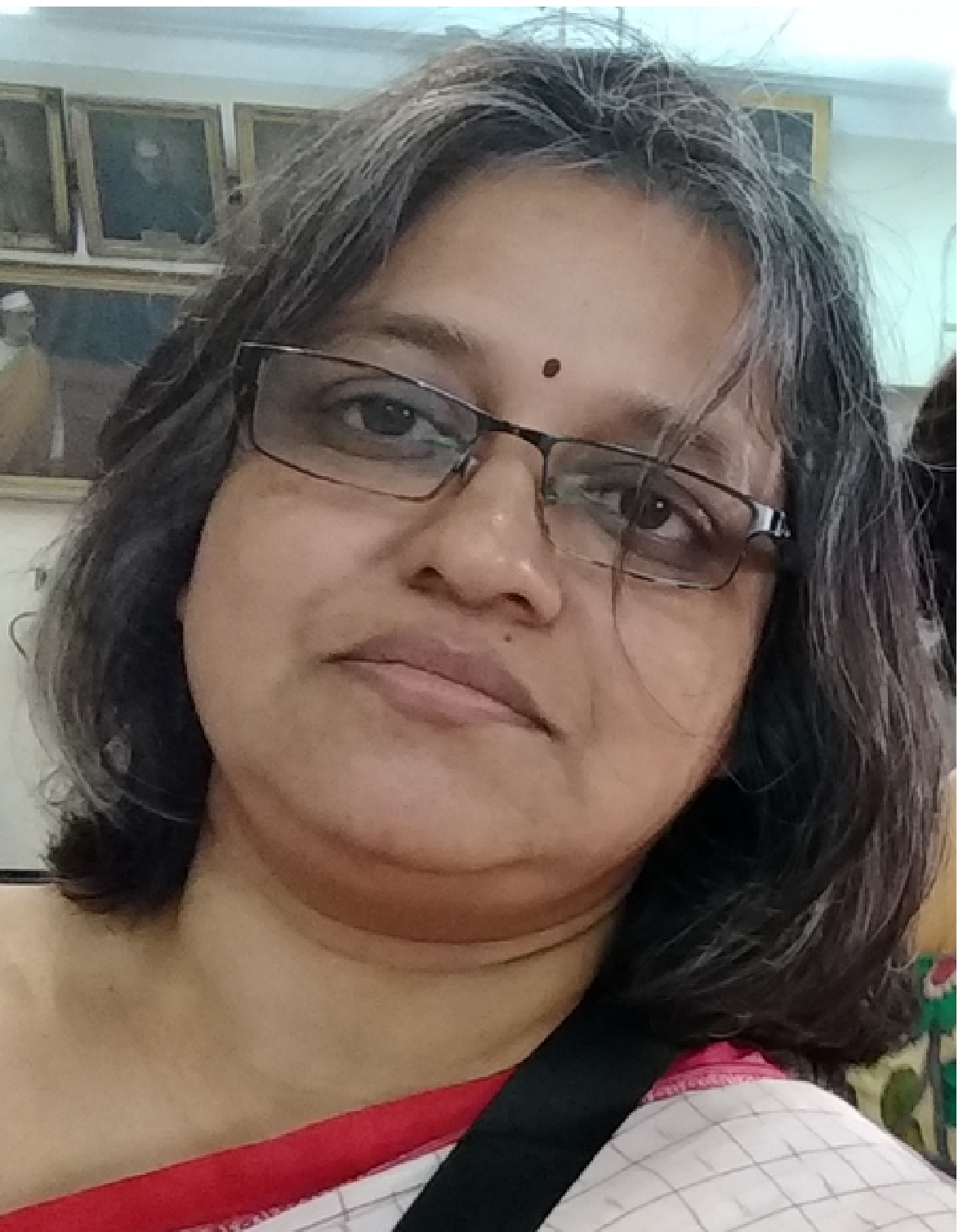}\\
 Sushan Konar  works  on  stellar  compact objects. She also writes popular
 science articles and maintains a weekly astrophysics-related blog called
 `{\em Monday Musings}'.}
\begin{abstract}
In  the previous  articles   of  this series,  we  have discussed  the
development  of musical  scales  particularly that  of the  heptatonic
scale which forms the basis of Western classical music today.  In this
last article, we take a look at  the basic structure of scales used in
Indian classical  music and  how different  {\em raga}s  are generated
through the simple process of scale shifting.
\end{abstract}
\monthyear{October 2019}
\artNature{SERIES  ARTICLE}

\section*{Introduction}
In a  certain generation many  people, all around the  world, received
their first lesson  in Western musical scales from  Julie Andrews when
she and the  von Trapp children sang `do-re-mi..'  in  `{\em The Sound
  of  Music}'.  In  India,  what  surprised  the  uninitiated  is  the
equivalence of this scale with  the `saptak' (a scale containing seven
basic notes) that forms the basis of Indian traditional music.
  

Indian classical  music is  a genre  that is  prevalent in  the Indian
sub-continent  and parts  of the  far-eastern reaches  of South  Asia.
There exist two  major traditions - the North  Indian tradition called
the {\em Hindustani} classical, and  the South Indian variant known as
the {\em  Carnatic} classical.  They began as  one but  later diverged
into   two    separate   forms    because   of    various   historical
reasons. However,  much of  the basic structure  remain the  same till
date.

The  guiding principle  of Indian  classical music  is to  exploit the
freedom accorded by the special nature of human sensitivity (discussed
in  article-I) to  the acoustic  frequencies.  \rightHighlight{A  {\em
    Raga} is  built from  a basic  scale called  a {\em  thaat}.}  The
primary characteristic of this genre is that it is based on a standard
set of melodic forms ({\em raga}s),  which are themselves built from a
basic set of  scales ({\em thaat}).  The {\em  raga}s basically define
the  overall mood  of the  music by  specifying scales  (ascending and
descending, which may or may not  be the same) and provide the general
prescription according to which a piece of music should be composed or
performed.  As  there is  no rigidity  about a set  piece of  music, a
musician is entirely  free to bring her/his individual  flavour to the
composition as long as the  prescription specific to a particular {\em
  raga} is adhered to.

\section{Basic Structure}

\begin{table}
\caption{Correspondence  between  the  Indian {\em  shruti}s  and  the
  Western notes.  Note that the {\em shuddha swara}s coincide with the
  pure notes of C-major. This is  because the Indian base note $s$ has
  been  matched  to  the  Western  $C$  and  Indian  {\em  saptak}  is
  intrinsically  a  major scale.  The  absolute  frequencies has  been
  obtained by setting $A$ to 440~Hz.}
\label{t-shruti}
%
\begin{tabular}{|l|l|l|l|l|} \hline
%
$Shruti$  	 &  ratio    &  $\nu$ (Hz) &  Note & $\nu$ (Hz) \\ \hline
{\em Chandovati} (\textcolor{magenta}{$sa$})  &  1 	  &  261.6256  & \textcolor{magenta}{C} & 261.6256 \\
{\em Dayavati} 	                              &  256/243  &  275.6220  & C\#                    & 277.1826 \\
{\em Ranjani} 	                              &  16/15    &  279.0673  &                        & \\
{\em Ratika} 	                              &  10/9     &  290.6951  &                        & \\
{\em Raudri} (\textcolor{magenta}{$re$})      &  9/8      &  294.3288  & \textcolor{magenta}{D} & 293.6648 \\
{\em Krodha} 	                              &  32/27    &  310.0747  & D\#                    & 311.1270 \\
{\em Vajrika} 	                              &  6/5      &  313.9507  &                        & \\
{\em Prasarini} (\textcolor{magenta}{$ga$})   &  5/4      &  327.0319  & \textcolor{magenta}{E} & 329.6275 \\
{\em Marjani}  (\textcolor{magenta}{$ma$})    &  4/3      &  348.8341  & \textcolor{magenta}{F} & 349.2282 \\
{\em Rakta} 	                              &  45/32    &  367.9109  & F\#                    & 369.9944 \\
{\em Sandipani} 	                      &  729/512  &  372.5098  &                        & \\
{\em Alapini} (\textcolor{magenta}{$pa$})     &  3/2      &  392.4383  & \textcolor{magenta}{G} & 391.9954 \\
{\em Madantī} 	                              &  128/81   &  413.4330  & G\#                    & 415.3047 \\
{\em Rohini} 	                              &  8/5      &  418.6009  &                        & \\
{\em Ramya} (\textcolor{magenta}{$dha$})      &  5/3      &  436.0426  & \textcolor{magenta}{A} & 440.0000 \\
{\em Ugra} 	                              &  27/16    &  441.4931  &                        & \\
{\em Ksobhini} 	                              &  16/9     &  465.1121  & A\#                    & 466.1638 \\
{\em Tivra} 	                              &  9/5      &  470.9260  &                        & \\
{\em Kumudvati} (\textcolor{magenta}{$ni$})   &  15/8     &  490.5479  & \textcolor{magenta}{B} & 493.8833 \\
{\em Manda} 	                              &  243/128  &  496.6798  &                        & \\
{\em Chandovati} (\textcolor{magenta}{$sa'$}) &  2        &  523.2511  & \textcolor{magenta}{C} & 523.2511 \\ \hline
\end{tabular}
%
\end{table}

In Indian  music there are 7  pure notes ({\em shuddha  swara}) - $sa$
($shadaj$), $re$ ($rishabh$), $ga$ ($gandhar$), $ma$ ($madhyam$), $pa$
($pancham$), $dha$ ($dhaiwat$) and $ni$ ($nishad$).  The first and the
fifth  notes $sa$  and $pa$  have fixed  frequencies and  are commonly
known as {\em  atal swara}s (invariant notes).  The other  5 notes are
variables and  the variants are known  as the {\em vikrita  swara}s or
the impure  notes. These  impure notes are  $\mathcal{R}, \mathcal{G},
\mathcal{D},  \mathcal{N}, \mathcal{M}$  corresponding to  the $komal$
(flat or lower  frequency) variants of $re, ga, dha,  ni$ and $teevra$
(sharp   or   higher   frequency)  variant   of   $ma$   respectively.
\leftHighlight{The Indian  {\em saptak}  roughly   corresponds  to the
  {\bf  Major} scale  of Western  tradition, and  consists of  12 {\em
    swara} - 7 {\em shuddha} and  5 {\em vikrita}.} An octave consists
of the seven  pure notes and is  known as a $saptak$,  the eighth note
having twice  the frequency of the  first note.  In reality  though, a
$saptak$ contains 12 notes - 7 pure and 5 impure. The seven pure notes
are obtained according to the ratio - 1, 9/8, 5/4, 4/3, 3/2, 5/3, 15/8
between   the  consecutive   notes.   From   our  earlier   discussion
(article-II) it is  easy to see that this corresponds  very closely to
the {\bf  Major} scale of the  Western tradition, as can  be seen from
{\em Table}~[\ref{t-shruti}],  though the temperament used  is neither
Pythagorean nor the ETS, but the {\em Just}.

\begin{table}
\caption{Indian {\em  saptak} - {\em mandra-saptak}  notes are denoted
  with $hasanta$  symbols ($_{_\smallsetminus}$) and  {\em tar-saptak}
  notes  are denoted  with  $ref$ ($'$)  symbols.  (There exist  other
  styles   of  notation   to  distinguish   the  notes   in  different
  $saptak$s.)}
\label{t-saptak}
%
  \begin{tabular}{|l|c|c|c|c|} \hline
Key  & Note & $\nu$ (Hz) & \multicolumn{2}{|c|}{\bf \em Saptak} \\ \hline
\textcolor{light}{W} & C$_3$                      & \textcolor{blue}{130.81}    & $sa_{_\smallsetminus}$            &  \\
{\bf B}              &                            & 138.59                      & $\mathcal{R}_{_\smallsetminus}$   &  \\ 
\textcolor{light}{W} & D$_3$                      & \textcolor{blue}{146.83}    & $re_{_\smallsetminus}$            &  \\ 
{\bf B}              &                            & 155.56                      & $\mathcal{G}_{_\smallsetminus}$   &  \\ 
\textcolor{light}{W} & E$_3$                      & \textcolor{blue}{164.81}    & $ga_{_\smallsetminus}$            &  \\ 
\textcolor{light}{W} & F$_3$                      & \textcolor{blue}{174.61}    & $ma_{_\smallsetminus}$            &  {\bf \em Mandra} \\ 
{\bf B}              &                            & 185.00                      & $\mathcal{M}_{_\smallsetminus}$   &  \\ 
\textcolor{light}{W} & G$_3$                      & \textcolor{blue}{196.00}    & $pa_{_\smallsetminus}$            & \\ 
{\bf B}              &                            & 207.65                      & $\mathcal{D}_{_\smallsetminus}$   & \\ 
\textcolor{light}{W} & A$_3$                      & \textcolor{blue}{220.00}    & $dha_{_\smallsetminus}$           & \\ 
{\bf B}              &                            & 233.08                      & $\mathcal{N}_{_\smallsetminus}$   & \\ 
\textcolor{light}{W} & B$_3$                      & \textcolor{blue}{246.94}    & $ni_{_\smallsetminus}$            & \\  \cline{4-5}
\textcolor{light}{W} & C$_4$                      & \textcolor{blue}{261.63}    & $sa$                        & \\ 
{\bf B}              &                            & 277.18                      & $R$                         & \\ 
\textcolor{light}{W} & D$_4$                      & \textcolor{blue}{293.67}    & $re$                        & \\ 
{\bf B}              &                            & 311.13                      & $\mathcal{G}$               & \\ 
\textcolor{light}{W} & E$_4$                      & \textcolor{blue}{329.63}    & $ga$                        & \\ 
\textcolor{light}{W} & F$_4$                      & \textcolor{blue}{349.23}    & $ma$                        & {\bf \em Madhya} \\ 
{\bf B}              &                            & 369.99                      & $\mathcal{M}$               & \\ 
\textcolor{light}{W} & G$_4$                      & \textcolor{blue}{392.00}    & $pa$                        & \\ 
{\bf B}              &                            & 415.30                      & $\mathcal{D}$               & \\ 
\textcolor{light}{W} & \textcolor{magenta}{A$_4$} & \textcolor{magenta}{440.00} & $dha$                       & \\ 
{\bf B}              &                            & 466.16                      & $\mathcal{N}$               & \\ 
\textcolor{light}{W} & B$_4$                      & \textcolor{blue}{493.88}    & $ni$                        & \\ \cline{4-5}
\textcolor{light}{W} & C$_5$                      & \textcolor{blue}{523.25}    & $sa'$                       & \\ 
{\bf B}              &                            & 554.37                      & $R'$                        & \\ 
\textcolor{light}{W} & D$_5$                      & \textcolor{blue}{587.33}    & $re'$                       & \\ 
{\bf B}              &                            & 622.25                      & $\mathcal{G}'$              & \\ 
\textcolor{light}{W} & E$_5$                      & \textcolor{blue}{659.26}    & $ga'$                       & \\ 
\textcolor{light}{W} & F$_5$                      & \textcolor{blue}{698.46}    & $ma'$                       & {\bf \em Tar} \\ 
{\bf B}              &                            & 739.99                      & $\mathcal{M}'$              & \\ 
\textcolor{light}{W} & G$_5$                      & \textcolor{blue}{783.99}    & $pa'$                       & \\ 
{\bf B}              &                            & 830.61                      & $\mathcal{D}'$              & \\ 
\textcolor{light}{W} & A$_5$                      & \textcolor{blue}{880.00}    & $dha'$                      & \\ 
{\bf B}              &                            & 932.33                      & $\mathcal{N}'$              & \\ 
\textcolor{light}{W} & B$_5$                      & \textcolor{blue}{987.77}    & $ni'$                       & \\ \hline 
\end{tabular}
%
%
\end{table}

In traditional  Indian music  a total of  22 micro-tones  or $shruti$s
were in  use instead of  the 12  tones discussed above.   The practice
continues to be  the same in the South-Indian  (Carnatic) music though
the  North-Indian (Hindustani)  system  is now  more  or less  12-tone
based. The division of the $saptak$  in 22 $shruti$s exploits the fact
that there exists a minimum interval  (in pitch or frequency) that can
be distinguished by human ear.   (Theoretically, an infinite number of
$shruti$s are possible but any  practical division would depend on the
actual size of the frequency interval that a listener can discern or a
musician can produce.)   The list of $shruti$s has been  shown in {\em
  Table}~[\ref{t-shruti}],  along  with  the   pure  notes  and  their
correspondence with the Western scale.  It  can also be seen from {\em
  Table}~[\ref{t-shruti}] that the  difference between the frequencies
of notes in  the Indian system and the  Equal-Tempered-Scale (ETS) are
rather small.  In fact, with  the introduction of the reed instruments
(piano, harmonium etc.)  in the Indian music scene (in particular, the
huge popularity of harmonium across musical genres) the difference has
all but disappeared.  Therefore, for  the sake of convenience we shall
use the  ETS even while talking  about the Indian scales  and notes in
this article.

As has  been mentioned  before, a $saptak$  corresponds to  an octave.
Three main $saptak$s are used  in Indian music.  Unlike Western music,
which has  an absolute frame  of reference, the Indian  system changes
from instrument  to instrument.   \rightHighlight{The three  main {\em
    saptak}s of the Indian tradition  are the {\em mandra, madhya} and
  {\em  tar saptak}.}  The middle  register, referred  to as  the {\em
  madhya saptak},  uses a  base note  that is  most comfortable  for a
particular musician (vocal or instrument); everything else is reckoned
from here.  The octave above this base  is referred to as the {\em tar
  saptak}; and  the lower  one is  known as  the {\em  mandra saptak}.
Additionally,  two octaves  above the  middle is  called {\em  ati-tar
  saptak}; three octaves above is  called {\em ati-ati-tar saptak} and
so  on.  The  reed instruments  also allow  us to  connect the  Indian
$saptak$s with the corresponding octaves of  an ETS in an easy manner,
as shown in {\em Table}~[\ref{t-saptak}].  It is also clearly seen how
the  Indian scale  corresponds to  the `major'  scale, since  the pure
tones of a $saptak$ follows the `T T S T T T S' pattern.

\section{Shifting the Scale}
One of  the main characteristic differences  between Western classical
music and Indian is in their approach to fixing the {\em tonic} or the
{\em base  note}. In  Western tradition,  as we  have seen  earlier, a
particular piece  of music  is set  for a  particular scale  (the home
octave, inclusive of  all the notes) and the instruments  are tuned to
play those specific frequencies.  On  the other hand, Indian music is,
more or less, independent of the  chosen home octave.  A performer can
choose  the base  note  ($sa$) of  the {\em  madhya  saptak} (or  more
precisely,  the  home octave)  according  to  her/his convenience  and
therefore  effectively  has infinite  freedom  in  doing so.   Indeed,
traditional  Indian  music makes  use  of  the infinite  possibilities
accorded by the  frequency continuum.  This freedom is  enjoyed by the
vocalists  and also,  to some  extent, by  the musicians  using string
instruments.  However,  for reed  instruments the  change of  the home
octave  would necessarily  be  discrete.  In  the  following we  shall
discuss two  different cases of  this {\em  shift} (both based  on the
discrete ETS) commonly made use of in Indian music.

\begin{table}
\vspace{-1.15cm}
\begin{tabular}{|l|c|c|c|c|c|} \hline
Note        & $\nu$                         & C                      &  F                           & A                            & B$_{\rm Flat}$       \\ 
            &  Hz                           &   Major                &    Major                     &   Major                      &  Major \\ \hline
C$_2$        & 65.406                       & $sa_{_\smallsetminus}$                   &                              &                              & \\ 
             & 69.296                       & $\mathcal{R}_{_\smallsetminus}$          &                              &                              & \\ 
D$_2$        & 73.416                       & $re_{_\smallsetminus}$                   &                              &                              & \\ 
             & 77.782                       & $\mathcal{G}_{_\smallsetminus}$          &                              &                              & \\ 
E$_2$        & 82.407                       & $ga_{_\smallsetminus}$                   &                              &                              & \\ 
F$_2$        & \textcolor{blue}{\bf 87.307} & $ma_{_\smallsetminus}$                   & \textcolor{blue}{\bf \em sa} &                              & \\ 
             & 92.499                       & $\mathcal{M}_{_\smallsetminus}$          & $\mathcal{R}$                &                              & \\ 
G$_2$        & 97.999                       & $pa_{_\smallsetminus}$                   & $re$                         &                              & \\ 
             & 103.83                       & $\mathcal{D}_{_\smallsetminus}$          & $\mathcal{G}$                &                              & \\ 
A$_2$        & \textcolor{blue}{\bf 110.00} & $dha_{_\smallsetminus}$                  & $ga$                         & \textcolor{blue}{\bf \em sa} & \\ 
B$_{\rm flat}$ & \textcolor{blue}{\bf 116.54} & $\mathcal{N}_{_\smallsetminus}$          & $\mathcal{M}$                & $\mathcal{R}$                & \textcolor{blue}{\bf \em sa} \\ 
B$_2$        & 123.47                       & $ni_{_\smallsetminus}$                   & $ma$                         & $re$                         & $\mathcal{R}$ \\ 
C$_3$        & 130.81                       & $sa$                               & $pa$                         & $\mathcal{G}$                & $re$ \\ 
             & 138.59                       & $R$                                & $\mathcal{D}$                & $ga$                         & $\mathcal{G}$ \\ 
D$_3$        & 146.83                       & $re$                               & $dha$                        & $ma$                         & $ga$ \\ 
             & 155.56                       & $\mathcal{G}$                      & $\mathcal{N}$                & $\mathcal{M}$                & $ma$ \\ 
E$_3$        & 164.81                       & $ga$                               & $ni$                         & $pa$                         & $\mathcal{M}$ \\ 
F$_3$        & 174.61                       & $ma$                               & $sa'$                        & $\mathcal{D}$                & $pa$ \\ 
             & 185.00                       & $\mathcal{M}$                      &                              & $dha$                        & $\mathcal{D}$ \\ 
G$_3$        & 196.00                       & $pa$                               &                              & $\mathcal{N}$                & $dha$ \\ 
             & 207.65                       & $\mathcal{D}$                      &                              & $ni$                         & $\mathcal{N}$ \\ 
A$_3$        & 220.00                       & $dha$                              &                              & $sa'$                        & $ni$ \\ 
             & 233.08                       & $\mathcal{N}$                      &                              &                              & $sa'$ \\ 
B$_3$        & 246.94                       & $ni$                               &                              &                              & \\ 
C$_4$        & 261.63                       & $sa'$                              &                              &                              & \\  \hline
\end{tabular}
%
\caption{Illustration of  {\em scale  change} - When  the base  note is
  moved  (by a  particular multiplicative  factor) keeping  the octave
  structure  intact, it  shifts  all  the notes  in  the entire  scale
  exactly by the same multiplicative factor.}
\label{t-shift}
\end{table}

\subsection{Scale  Change}
The simple {\em shift} of the home octave (popularly known as the {\em
  \bf scale  change} in India) is  just that.  The change  of the base
note from one  frequency to another one, keeping the  structure of the
{\em saptak} intact.  \rightHighlight{A scale-change is a simple shift
  in the  base frequency, without any  change in the structure  of the
  music.} Remember,  we have twelve  {\em swara}s to the  {\em saptak}
corresponding  to the  twelve  steps of  an ETS,  the  frequency of  a
particular {\em  swara} being  $2^{\frac{1}{12}}$ higher than  the one
immediately preceding it.  {\em Table}~[\ref{t-shift}] illustrates the
{\em scale change}  for a reed instrument.   For vocalists accompanied
by  harmonium, the  scales  spanning $G_2-B_2$  are  quite popular  in
modern  Indian music.  Typically,  male voices  prefer  scales with  a
higher  frequency base  note  compared to  those  preferred by  female
voices,  signifying  the natural  pitch(frequency)-difference  between
male  and female  voices.   Of course,  there exist  a  huge range  of
natural  frequencies at  which a  particular vocalist  is comfortable.
For example,  there are people  who feel  most comfortable to  sing at
$F_2$ implying that  the natural frequency of their voice  is a factor
of $2^{\frac{6}{12}}$ ($\simeq$1.414) lower than the natural frequency
of someone singing at $B_2$.
\begin{table}[h]
%
\begin{tabular}{|l|c|c|c|c|c|c|c|c|c|c|c|c|c|c|} \hline
$Bilawal$   &  $sa$ & $re$ & $ga$ & $ma$ & $pa$ & $dha$ & $ni$  & $sa'$ & *     & *     & *     & *     & *    & * \\ \hline
$Kafi$ 	     &   *   & $sa$ & $re$ & $ga$ & $ma$ & $pa$  & $dha$ & $ni$  & $sa'$ & *     & *     & *     & *    & * \\ \hline
$Bhairavi$   &   *   & *    & $sa$ & $re$ & $ga$ & $ma$  & $pa$  & $dha$ & $ni$  & $sa'$ & *     & *     & *    & * \\ \hline
$Kalyan$     &   *   & *    & *    & $sa$ & $re$ & $ga$  & $ma$  & $pa$  & $dha$ & $ni$  & $sa'$ & *     & *    & * \\ \hline
$Khammaj$     &   *   & *    & *    & *    & $sa$ & $re$  & $ga$  & $ma$  & $pa$  & $dha$ & $ni$  & $sa'$ & *    & * \\ \hline
$Asavari$     &   *   & *    & *    & *    & *    & $sa$  & $re$  & $ga$  & $ma$  & $pa$  & $dha$ & $ni$  & $sa'$ & * \\ \hline
----        &   *   & *    & *    & *    & *    & *     & $sa$  & $re$  & $ga$  & $ma$  & $pa$  & $dha$ & $ni$  & $sa'$ \\ \hline
\end{tabular}
%
\vspace{0.5cm}
\caption{Illustration of $Murchhana$. Only  $Hindustani$ names
  of the $raga$s are indicated. See text for an explanation.}
\label{t-murchana01}
\end{table}

Interestingly, likely due to its huge popularity across musical genres
in  India,  the harmonium  has  now  been  modified to  incorporate  a
scale-changing mechanism. Using  this, one can move from  a scale (say
$F$) to another (say $B$) without having to experience a change in the
piano  keys.  The harmonium  player  is  then  able  to use  the  same
fingering (that  s/he is used  to) for all  the shifts in  home octave
with the scale-changing mechanism of the instrument taking care of the
shift in frequency rather than a change in keystrokes.

\subsection{\boldmath $Murchhana$}
On the  other hand,  there is  a more complex,  and far  more profound
change of  scale that  is known  to Indian  classical music.   This is
known as  $murchhana$ in  Hindustani (North  Indian) classical  and as
$grahabhedam$   in  Carnatik   (South  Indian)   classical  tradition.
Evidently, the  Carnatik name gives  away the underlying  logic behind
this  -   $graha$  means  `position'  and   $bhedam$  means  `change'.
\rightHighlight{{\em  Murchhana}  or {\em  grahabhedam}  is  a way  of
  changing the  scale which alters  the basic scale structure  or {\em
    thaat}.} The process literally means  a change of position. Indeed
at a first  glance it appears to be  no more than a shift  of the base
note   and   all    the   subsequent   notes   as    shown   in   {\em
  Table}~[\ref{t-murchana01}]~\footnote{The   corresponding   Carnatik
  names  of  the {\em  raga}s  are  - Dhirashankarabaranam  (Bilawal),
  Kharaharapriya   (Kafi),    Hanumantodi   (Bhairavi),   Mechakalyani
  (Kalyan), Harikamboji (Khammaj), Natabhairavi (Asavari).}.  However,
this looks  completely counter-intuitive. We  have just seen  that the
Indian  music allows  for  any  shift of  the  home  octave.  By  that
argument, this  shift is not likely  to produce anything new.   On the
other hand, we also know that each  $raga$ has its own specific set of
$swara$s which gives it a  particular flavour specific to that $raga$.
Yet, we  are moving from one  particular $raga$ to another,  simply by
shifting  the $saptak$  by  a  number of  $swara$s  according to  this
prescription. Neither of these are satisfied if we consider this table
naively. Nor is it clear why the  last shift indicated in the table is
not an extant $raga$.

\begin{figure}
%
\hspace{-1.5cm}
  \includegraphics[width=13.5cm]{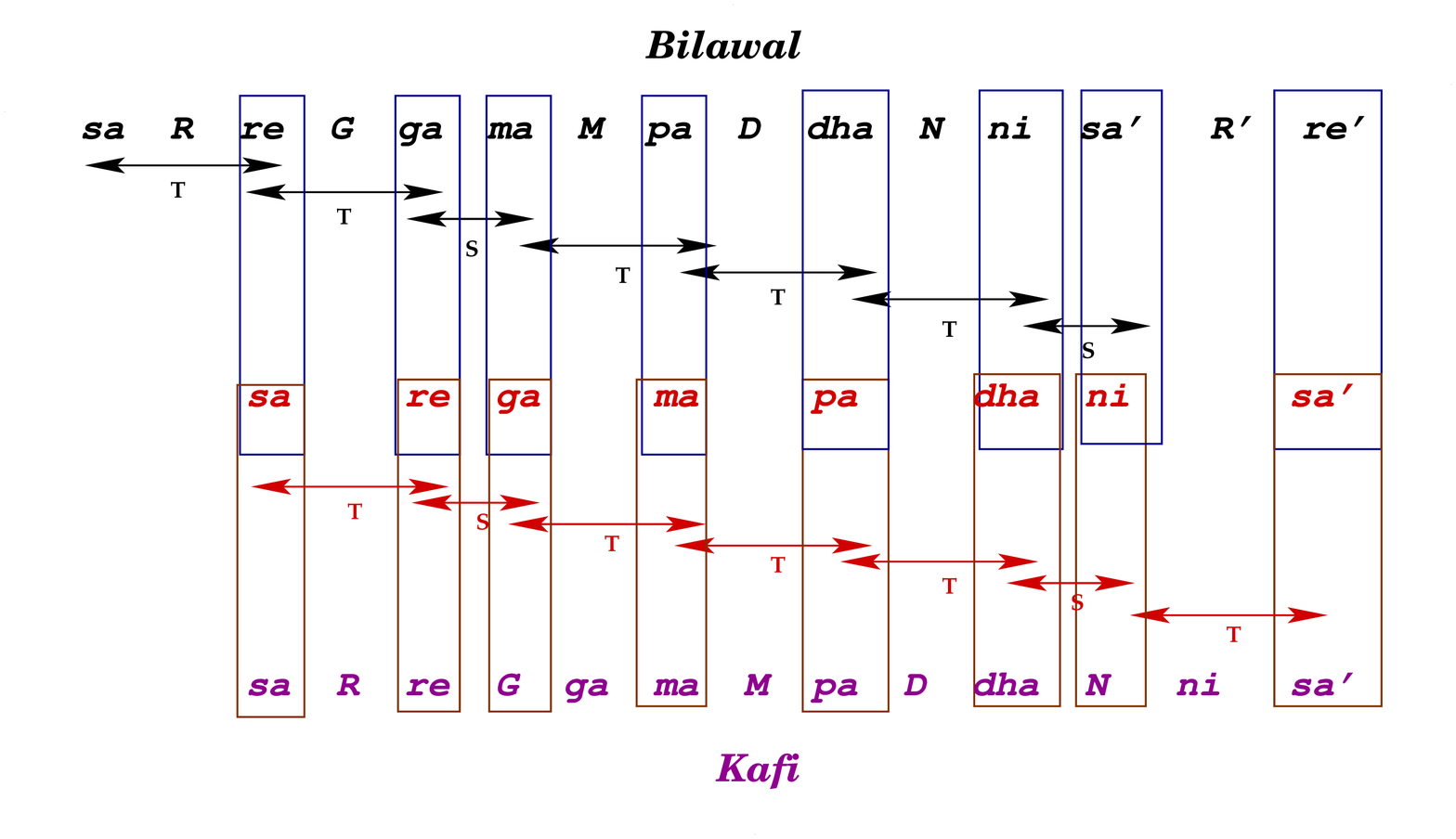}
%
\caption{Application of  the rules  of {\em  Murchhana} to  obtain {\em
    Raga Kafi} from {\em Raga Bilawal}.}
\label{f-murchana}
\vspace{-1.0cm}
\eef

Here, we  need to remember that  the {\em shuddha swara}s  or the pure
tones are not  equidistant (in a logarithmic sense) on  an ETS. But if
all the  12 tones ({\em  shuddha +  vikrita}) are taken  together then
they give  us 12 equidistant  notes. Whenever we map  a set of  7 {\em
  shuddha swara}s to another set of  7 {\em shuddha swara}s we are not
performing a  fixed frequency shift  (as was  the case earlier,  for a
simple `scale-change')  but something far more  complicated.  Consider
{\em raga Bilawal} which is characterised by a pure $saptak$, i.e., by
all  of  the  seven  {\em shuddha  swara}s.   Since,  Indian  $saptak$
corresponds to a  major scale of the Western tradition,  the notes are
separated by a  `T T S T T  T S' pattern, where $T$ stands  for a tone
(factor of $2^{\frac{2}{12}}$) and $S$  stands for a semi-tone (factor
of      $2^{\frac{1}{12}}$).      This      is     illustrated      in
Fig.[\ref{f-murchana}]. Now,  let us follow the  prescription given in
{\em Table}~[\ref{t-murchana01}] and  move the  $swara$s by one  position to
obtain {\em raga Kafi}. Note that the $swara$s of the shifted $saptak$
are  separated  by `T  S  T  T  T S  T'  pattern  (second row  of  the
figure). According to the Western definition,  it is no longer a major
scale.  However,  the Indian  $saptak$ adheres  strictly to  the major
scale. Therefore, if  we define a major scale or  a true $saptak$ with
the shifted base note ($sa$) we  obtain the pattern given by the third
row in Fig.[\ref{f-murchana}]. Comparing the  second and the third row
of the figure it is easy to  see that, instead of all the {\em shuddha
  swara}s of {\em  raga Bilawal} now we have two  {\em vikrita swara}s
(namely, $\mathcal{G}, \mathcal{N}$) for {\em raga Kafi}.

This is  how a  new $raga$  is created  by shifting  the base  note in
Indian  tradition  and  is  known  as  $murchhana$.   The  process  is
illustrated for the entire $saptak$ in Figure~[\ref{t-murchana02}]. It
needs to be noted that for all  the shifts, barring the last one, $pa$
remains  a {\em  shuddha swara}.  \ \rightHighlight{A  {\em murchhana}
  that shifts the  {\em atal swar `pa'} is not  allowed.} As it should
be, because in  Indian tradition $sa$ and $pa$ are  the two notes that
do  not have  any variant  (and $pa$  is actually  the `fifth'  of the
Pythagorean scale).  Since this condition is not satisfied in the last
shift ($pa$ is shifted to $\mathcal{M}$  here) it is not considered to
be a valid $raga$.

Of course,  Indian classical genre is  not confined to just  these six
{\em murchhana}s or  six {\em raga}s.  A multitude of  new {\em raga}s
can be created remembering that  traditional Indian scale consists not
seven  but, at  least 22  {\em  shruti}s (or  more). Also,  it is  not
mandatory to have  7 base notes, a {\em raga}  can also be constructed
with lesser  number of  base notes. Rather  complex theories  of music
exist (differing  significantly from one  side of the Vindhyas  to the
other) that deal  with the family of this large  number of extant {\em
  raga}s.  However, for the uninitiated, understanding this simple yet
elegant logic  underlying the {\bf  \em sounds  of music} of  the {\em
  raga}s appears to be a beautiful exercise in itself.

\begin{figure}
\vspace{-5.0cm}
\hspace{-5.0cm}
  \includegraphics[width=22.5cm,angle=-90]{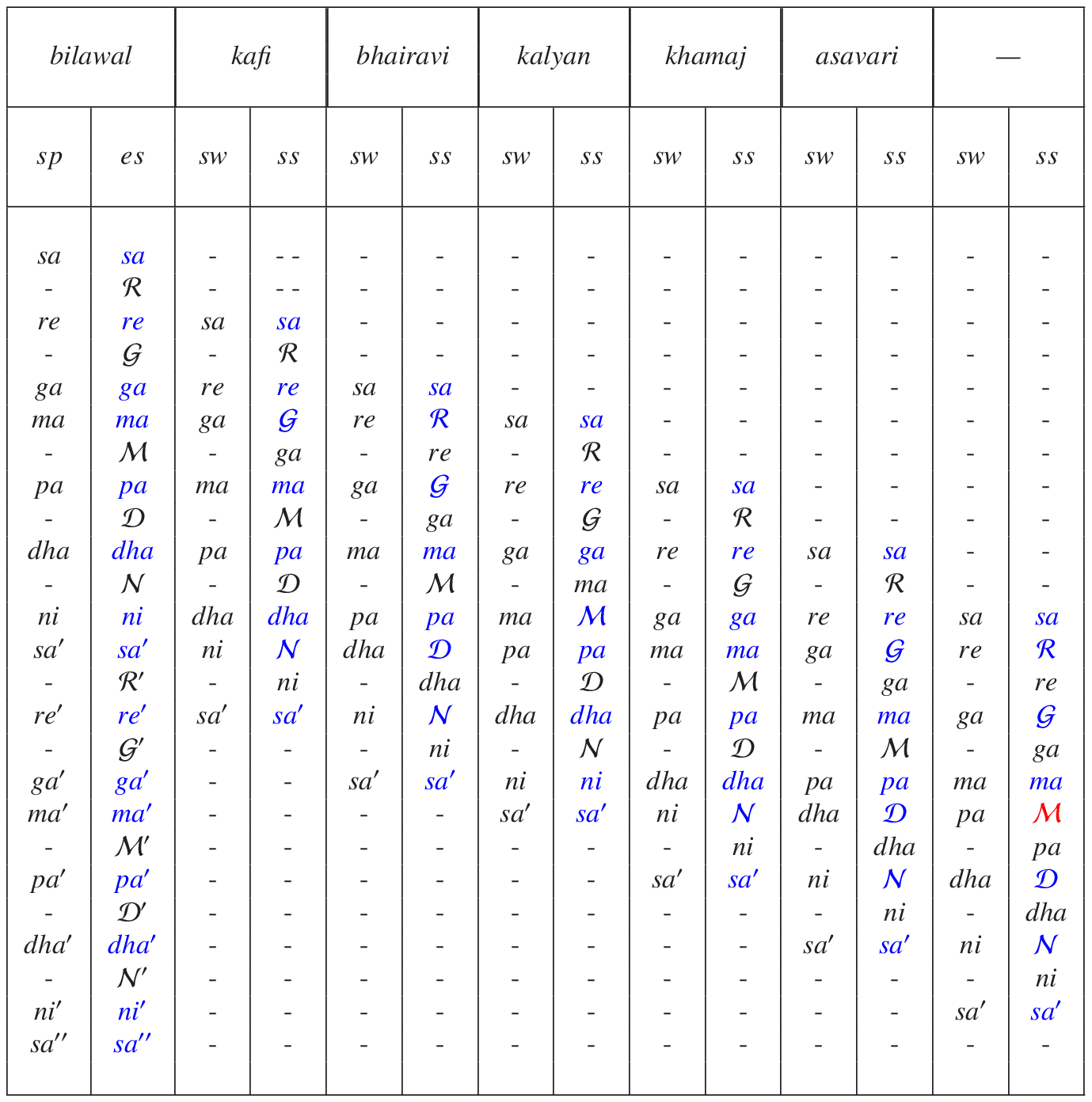}
\vspace{-3.5cm}
\caption{Application of {\em Murchhana} through the entire {\em saptak}.
Legends : {\em sp - saptak, es - extended saptak, sw - swara, ss - shifted saptak}}
\label{t-murchana02}
%
%
\eef

\section*{Acknowledgments}
Thanks  are  due to  the  music  exponents  of Surajhankar,  Pune  (in
particular, Ratnamanjari  Munshi, Sripurna Mitra, Uma  Roy \& Madhumita
Ghosh) for introducing me to the world of music. I am also indebted to
Sushruti Santhanam and Achintya  Prahlad for illuminating discussions;
and  to Ushasi  Roy Choudhury  for help  in fixing  some of  the scale
related issues.

\correspond{Sushan Konar \\
NCRA-TIFR \\
Pune 411007 \\
India \\
Email: sushan@ncra.tifr.res.in}

\end{document}